\documentclass[aps,prl,floatfix,onecolumn,showpacs]{revtex4}
\usepackage{graphicx,epsf}%
\newcommand {\apgt} {\ {\raise-.5ex\hbox{$\buildrel>\over\sim$}}\ }
\newcommand {\aplt} {\ {\raise-.5ex\hbox{$\buildrel<\over\sim$}}\ }
\usepackage{graphicx}
\usepackage{amssymb}

\begin{document}

\title{The Polar Nano Regions $\rightleftharpoons$ Relaxor Transition in 
$Pb_{1-X}(Sc_{1/2}Nb_{1/2})O_{3-X}$; $X=bulk~concentration~of~
nearest~neighbor~[Pb-O]~divacancies$.
}

\author{B. P. Burton}
\affiliation{Materials Science and Engineering Division,
Materials Measurement Laboratory,
National Institute of Standards and Technology
Gaithersburg, MD 20899-8520, USA}

\author{Eric Cockayne}
\affiliation{Materials Measurement Science Division,
Materials Measurement Laboratory,
National Institute of Standards and Technology
Gaithersburg, MD 20899-8520, USA}

\author{D. B. Gopman}
\affiliation{Materials Science and Engineering Division,
Materials Measurement Laboratory,
National Institute of Standards and Technology
Gaithersburg, MD 20899-8520, USA}

\author{Gunay Dogan}
\affiliation{Materials Science and Engineering Division,
Materials Measurement Laboratory,
National Institute of Standards and Technology
Gaithersburg, MD 20899-8520, USA}

\author{Sarah Hood}
\affiliation{Materials Science and Engineering Division,
Materials Measurement Laboratory,
National Institute of Standards and Technology
Gaithersburg, MD 20899-8520, USA}

\affiliation{Hood College, Frederick, MD 21701, USA}

\begin{abstract}

In previous work, molecular dynamics simulations based on a
first-principles-derived effective Hamiltonian for 
$Pb_{1-X}(Sc_{1/2}Nb_{1/2})O_{3-X}$~ (PSN), with nearest-neighbor Pb-O 
divacancy pairs, was used to calculate 
$X_{\rm [Pb-O]}$~vs.~T, phase diagrams for PSN with:
ideal rock-salt type chemical order; nanoscale chemical short-range order; 
and random chemical disorder.  Here, we show that the 
phase diagrams should include additional regions in which a 
glassy relaxor-phase (or state) is predicted. With respect to
phase diagram topology, these results strongly support the 
analogy between relaxors and magnetic spin-glass-systems.  
\\

\end{abstract}


\maketitle

\begin{center}
Submitted  for publication in Physical Review B, \today
\end{center}

\section{Introduction}

Heterovalent perovskite-based $Pb(B,B^{\prime})O_{3}$~
relaxor ferroelectrics (RFE) \cite{Smolensky, Cross}, such as
$Pb(Sc_{1/2},Nb_{1/2})O_{3}$~ (PSN), $Pb(Sc_{1/2},Ta_{1/2})O_3$~ (PST),
and $Pb(Zn_{1/3},Nb_{2/3})O_3$~ (PZN) 
and, $relaxors$~ [which have no ferroelectric (FE) ground-state] such as
$Pb(Mg_{1/3},Nb_{2/3})O_3$~ (PMN) and 
$Pb(Mg_{1/3},Ta_{2/3})O_3$~ (PMT),
are technologically important transducer/actuator
materials with extraordinary dielectric and electromechanical
properties.  Chemically disordered PSN exhibits 
polar nano-regions (PNR) characteristics (more polarizable PNR in a
less polarizable matrix) above a normal FE-transition at
$T_{FE} \approx$373~K.  Chu $et~al.$ \cite{Chu0}
demonstrated that the addition of 1.7 atomic~\% Pb-O 
divacancies depresses the FE transition temperature (T), 
from $T_{FE} \approx$373~K to $T_{FE} \approx$338~K, and broadens the 
T-range in which PNR properties, {\it e.g.} frequency dispersion in the
dielectric response, are observed.  
Chu $et~al.$~ also reported similar and more complete results for 
isostructural PST\cite{Chu1,Chu2,Chu3}.
These results suggest that a sufficient bulk concentration of 
divacancy pairs, $X_{\rm [Pb-O]}$, will drive the system to a  
relaxor ferroelectric (RFE) state, with an FE-ground-state, or
to a fully relaxor state, without an FE-ground-state, at 
$X_{C}~<~X_{\rm [Pb-O]}$, where $X_{C}$~ is the critical 
composition at which $T_{FE} \rightarrow 0K$. 

Chemical disorder and 
defects such as Pb-vacancies (V$_{Pb}$) \cite{Bellaiche07}, oxygen
vacancies (V$_{O}$) or charge-compensating nearest neighbor (nn) 
Pb-O divacancy pairs (V$_{Pb-O}^{nn}$) \cite{CockaynePbO},
are sources of local, $random~fields$ ($\vec{h_i}$)
e.g. \cite{Kleemann2006,Burton2006,Kleemann2012} 
(angle brackets indicate a simulation box average).
Hence, the T vs. $X_{\rm [Pb-O]}$~ phase diagrams presented here 
are topologically equivalent to the T vs. $\langle \vec{h_i} \rangle $~ 
diagrams that are typically drawn for analytical mean-field models of 
magnetic spin-glass (SG) systems \cite{Sherr1,Sherr2,Sherr3,Sherr4}.

Recent publications by Sherrington \cite{Sherr1,Sherr2,Sherr3,Sherr4}
emphasized an analogy between relaxor ferroelectrics and magnetic
SG with $soft-pdeudospins$; $i.e.$~ magnetic spins or ferroelectric 
displacements with variable magnitudes and arbitrary orientations.
Pseudospin-psuedospin interactions in these models are frustrated
(random-bond frustration \cite{EA1975}), and
the combination of frustration plus quenched chemical disorder \cite{QUENCH} 
are identified as essential constituents of relaxors.
The model used here: also has $soft~ pseudospins$~ ($\xi_{i}$) at each
Pb-site; first-, second-, and third-nn $\xi_{i}-\xi_{j}$-pairwise 
interactions, plus 4'th through 39'th-nn $\xi_{i}-\xi_{j}$-pair 
dipole-dipole interactions; and $\vec{h_i}$~ at each Pb-site.
An analysis of $\vec{h_i}$~ that is based on nn Pb--B-site pairs in 
an ideal perovskite structure with a random cation configuration
\cite{Burton2006} indicates a distribution
of orientations such that 34\% are along $<111>$-type directions;
21\% are $<001>$-type; 19\% are $<110>$-type; 19\% are $<113>$-type;
and 7\% are $<000>$\cite{Burton2006} (weighted by $\vec{h_i}$-strength
the corresponding percentages are: 29\% $<111>$, 21\% $<001>$, 
23\% $<110>$, and 27\% $<113>$). The $\vec{h_i}$~ used for
the calculations presented here were calculated as the local field 
imposed by the whole simulation box. 
In this model, $\xi_{i}-\xi_{j}$~ pairwise 
interactions are all FE in character, hence the $\vec{h_i}$~ and 
[Pb-O]-divacancies are the only sources of frustration; and 
ideally NaCl-ordered pure PSN is unfrustrated.

Results presented here require changes in
the phase diagrams that were presented in \cite{Burton2008}. 
The field that was formerly referred to as the RFE-region
in  $T(\rm X_{[Pb-O]})$~ vs. phase diagrams \cite{Burton2008} is
now divided into: 1) a PNR-region, in which spatially static but
orientationally dynamic PNR (centered on $\approx$2~nm diameter 
chemically ordered regions \cite{Burton2005})
are embedded in a less polarizable matrix; and 
2) an RFE/relaxor-region ($relaxor-region$~ for brevity) in which PNR 
have more static orientations. 
The T($X_{\rm [Pb-O]})$-curve [$i.e.$~ T($\langle \vec{h_i} \rangle $)-curve]
that divides the PNR-region from the relaxor-region
is referred to as $T^{\bigstar}(X_{\rm [Pb-O]})$. Dkhil \cite{Dkhil} referred to 
$T^{\bigstar}$~ as "...a local phase transition that gives rise to the 
appearance of static polar nanoclusters." 
We reject the phrase "local phase transition," because (strictly) 
phase transitions only occur in infinite systems, and because our results 
suggest a weakly first-order transition, however, we do predict
a subtle stiffening of PNR-orientations below $T^{\bigstar}$.

In previous simulations\cite{Burton2006,Burton2008}, the presence of V$_{Pb}$~
vacancies\cite{Bellaiche07} or V$_{Pb-O}^{nn}$~
divacancies\cite{Burton2006,Burton2008} in PSN 
lead to more diffuse FE phase transitions, with broadened dielectric
susceptibility peaks; however, the RFE/relaxor-phase (state?) was not 
clearly delineated.  Here, simulations are used to construct 
$X_{\rm [Pb-O]}$~vs.~T phase diagrams
for $Pb_{1-X}(Sc_{1/2}Nb_{1/2})O_{3-X}$~ with random, perfectly 
rock-salt ordered and nano-ordered (NO) 
cation configurations as in \cite{Burton2008}.
The NO configuration  has 20 NaCl-type ordered clusters in a percolating random
matrix.  Divacancy concentration- and T-ranges for PE- and
FE-phases, and for "RFE-states", were identified from changes in polarization
correlations\cite{Tinte06}, but the RFE/relaxor-phase $per~ se$~ was not 
delineated.

\section{Simulations}
\subsection{The Model Hamiltonian}
Simulations were performed using the first-principles based effective
Hamiltonian $H_{eff}$~ that is described in detail in~\cite{Burton2006};
it expands the potential energy of PSN in a Taylor series about a
high-symmetry perovskite reference structure, including those degrees
of freedom relevant to FE phase transitions:

\begin{eqnarray}
 H_{eff} &=& H(\{\vec{\xi}_{i}\})+H(e_{\alpha \beta})
+ H(\{\vec{\xi}_{i}\},e_{\alpha \beta})+PV \nonumber \\
&+& H(\{\vec{\xi}_{i}\},\{\sigma_{l}\},\{{\rm V_{Pb-O}}\}) 
\label{H:eq}
\end{eqnarray}

\noindent
where $\{\vec{\xi}_{i}\}$~ represents Pb-site centered local polar 
distortion variables of arbitrary magnitudes and orientations;
$e_{\alpha \beta}$~ is a homogeneous strain term; 
$H(\{\vec{\xi}_{i}\},e_{\alpha \beta})$~ is a strain coupling term;
and $PV$~ the standard pressure-volume term.  The first four
terms are sufficient to model pressure-dependent
phase transitions in a normal FE perovskite without
local fields~\cite{Rabe}.
The fifth term, $H(\{\vec{\xi}_{i}\},\{\sigma_{l}\},\{{\rm V_{Pb-O}}\})$,
represents coupling between polar variables and ``random" local 
fields, $\vec{h_{i}}$,~\cite{ISFD7,TMS,Burton2006} from:
1) screened electric fields from the quenched distribution of 
Sc$^{3+}$~ and Nb$^{5+}$~ ions $\{\sigma_{l}\}$; and 2) by 
V$_{[Pb-O]}$.

As described in \cite{Burton2008} all simulations were done 
with a $40 \times 40 \times 40$~ MD-supercell, 
in which each Pb-atom is associated with a local 
distortion vector, $\vec{\xi}_{i}$, that indicates
the displacement of lead atom $Pb_{i}$~ from its ideal perovskite position.
The effective Hamiltonian in Eqn. \ref{H:eq} was used to 
derive equations of motion, with an MD time-step of 0.06 picoseconds.

Divacancies are modeled by replacing $40^3X_{\rm [Pb-O]}$~ 
randomly selected local distortion variables with fixed dipole moments 
corresponding to V$_{Pb-O}^{nn}$~ divacancy pairs ($i.e.$~ local fields directed,
from a Pb-site, along one of the 12 $\langle$110$\rangle$-type vectors).

\subsection{Order Parameters}
Curves for the Burns temperatures, $T_{B}(X_{\rm [Pb-O]})$, \cite{Burns83} and the 
FE-transitions, $T_{FE}(X_{\rm [Pb-O]})$~
are identical to those in \cite{Burton2008}. Curves for 
$T^{\bigstar}(X_{\rm [Pb-O]})$~
were located by plotting T-dependent $q_{\xi\xi}$- and $q_{\Delta t}$-curves 
where: $q_{\xi\xi}$~ is the self-overlap order
parameter, \cite{Castellani2005} Eqn. \ref{xixi:eq}; 
and $q_{\Delta t}$~ Eqn. \ref{QDt:eq}, 
is an autocorrelation function that compares the displacement of 
atom $\xi_{i}$~ at time-$t$~ with $\xi_{i}$~ at time-$t+\Delta t$ (typically, 
$\Delta t$~ = 100~MD-snapshots = 6.0 picoseconds). 

The idea behind $q_{\Delta t}$~ is that
a time-sensitive order parameter may be more sensitive to the sort
of PNR-stiffening referred to by Dkhil: \cite{Dkhil}

\begin{eqnarray}
 q_{\xi\xi} = \frac{1}{N}\sum_{i}\langle\vec{\xi}_{i}\cdot\vec{\xi}_{i}\rangle 
\label{xixi:eq}
\end{eqnarray}

and 

\begin{eqnarray}
 q_{\Delta t} = \frac{1}{N}\sum_{i}\langle\vec{\xi}_{i,t}\cdot\vec{\xi}_{i,t+100}\rangle
\label{QDt:eq}
\end{eqnarray}

where: N is the number of Pb-sites; summations are over the all Pb-displacements;
and the averaging represented by angle brackets is over the 
last 1000 MD-snapshots in a 3000- or 5000 snapshot series (see below). 
Within the precision of these simulations, both order parameters yield the same
results for $T^{\bigstar}(X_{\rm [Pb-O]})$. 

Numerical simulations can not distinguish between crossovers and 
phase transitions where: crossovers correspond to inflection points in 
$q_{\xi\xi}(T)$~ and/or $q_{\Delta t}(T)$; and phase 
transitions correspond to discontinuities in first- or second-T-derivatives 
of $q_{\xi\xi}(T)$~ and/or $q_{\Delta t}(T)$~ ($i.e.$~ first-order, 
or continuous- or critical-transition, respectively \cite{Fisher}).  
Because the results for random- and 
NO-cation configurations strongly suggest a (weakly) first-order phase transition, 
$T^{\bigstar}$~ will be referred to as a phase $transition$, and the relaxor will
be referred to as a $phase$, but with the caveat that $T^{\bigstar}$~ may 
actually mark a crossover, and the relaxor would then be a $state$. 

Order parameter values were calculated from MD-snapshots that were taken
every 100 MD time-steps in a series of 3000 or 5000 
MD-snapshots (enough snapshots that $q_{\xi\xi}(T)$~ and $q_{\Delta t}(T)$,
are approximately constant for 1000 snapshots); 
3000 for the NaCl-ordered and random cation configurations; 
5000 for the NO configuration. 
Plotted order-parameter values are averages
over the last 1000 MD snapshots in a series.

\section{Results}

\begin{figure} [!htbp]
\includegraphics[width=80mm]{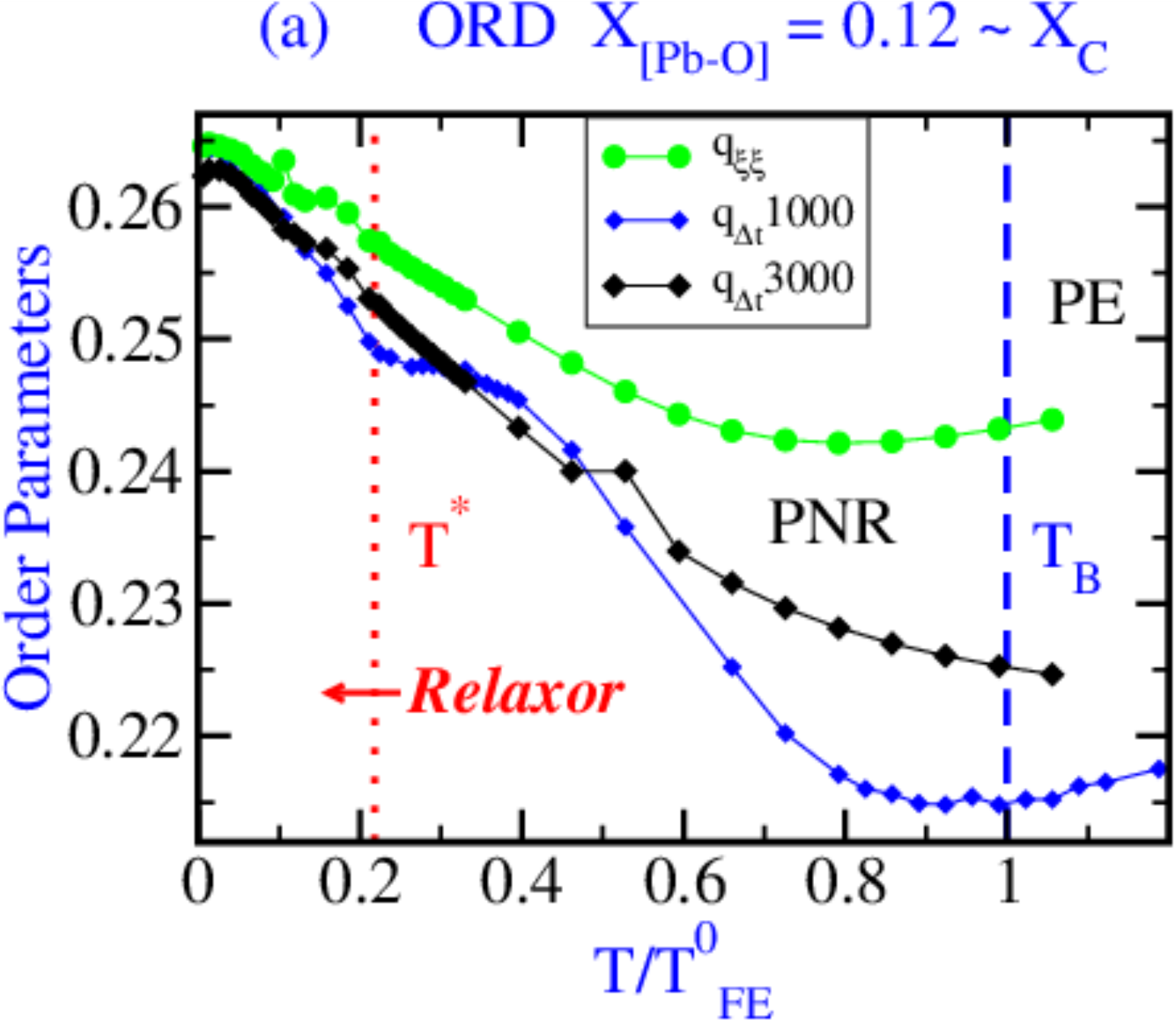}
\hspace{0.5cm}
\includegraphics[width=80mm]{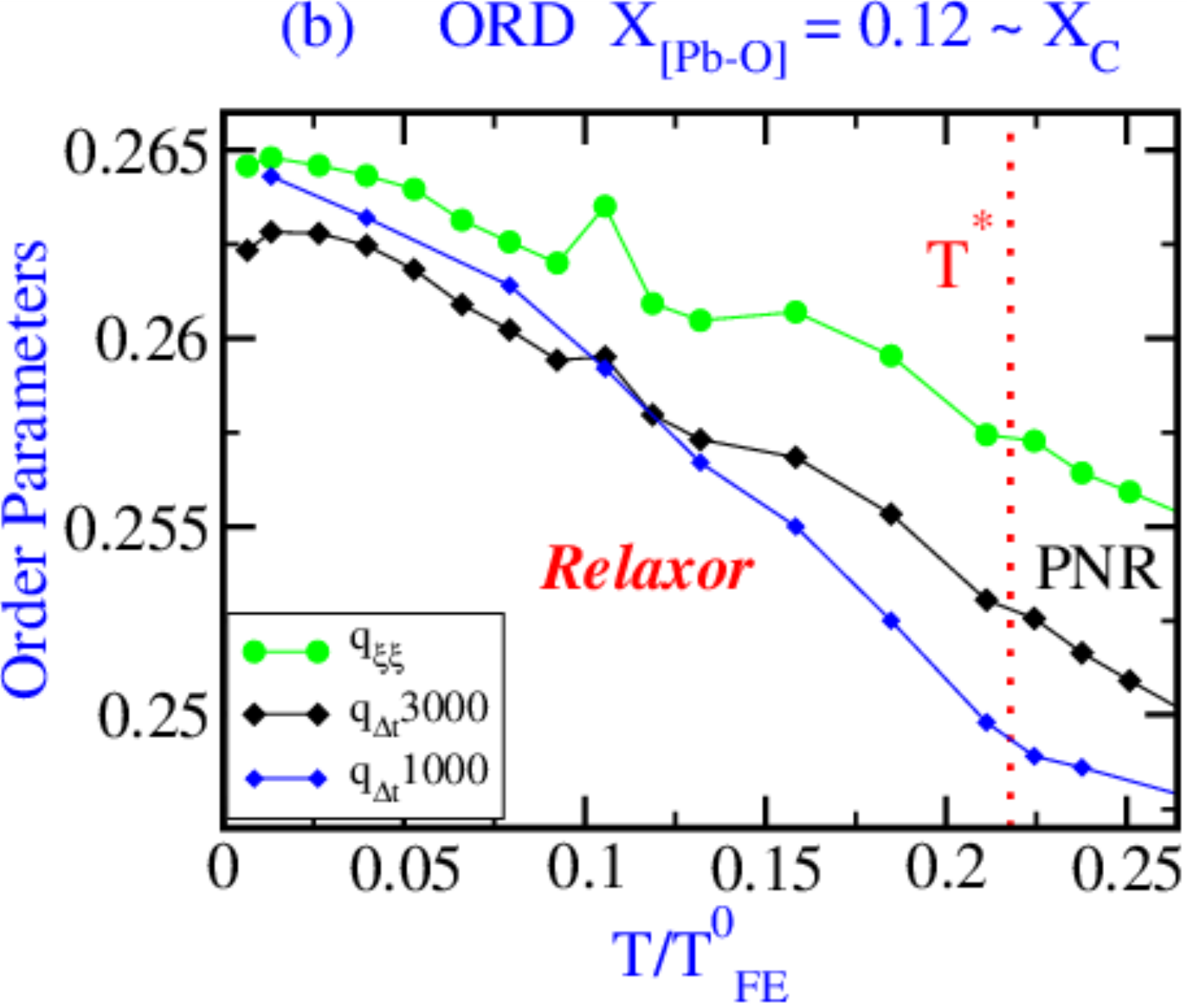}
\caption{Order parameters that were used to define the
relaxor-region in $Pb_{1-X}(Sc_{1/2}Nb_{1/2})O_{3-X}$, with
ideal rock-salt type Sc:Nb-chemical order: 
$q_{\xi\xi}(T)$~ is the self-overlap order parameter (Eqn. \ref{xixi:eq}); and 
$q_{\Delta t}$~ (Eqn. \ref{QDt:eq}) is a temporal autocorrelation function
($q_{\Delta t}$1000 and $q_{\Delta t}$3000 are results from
1000- and 3000-snapshots, respectively).  
Panels: (a) is the full diagram; (b) is an enlargement of the low-T portion 
of the diagram. Here, $T^{\bigstar}$~ looks as though it may mark a continuous 
transition, or a crossover.
}
\label{SSORD:fig}
\end{figure}

\begin{figure} [!htbp]
\vspace{1.5cm}
\includegraphics[width=80mm]{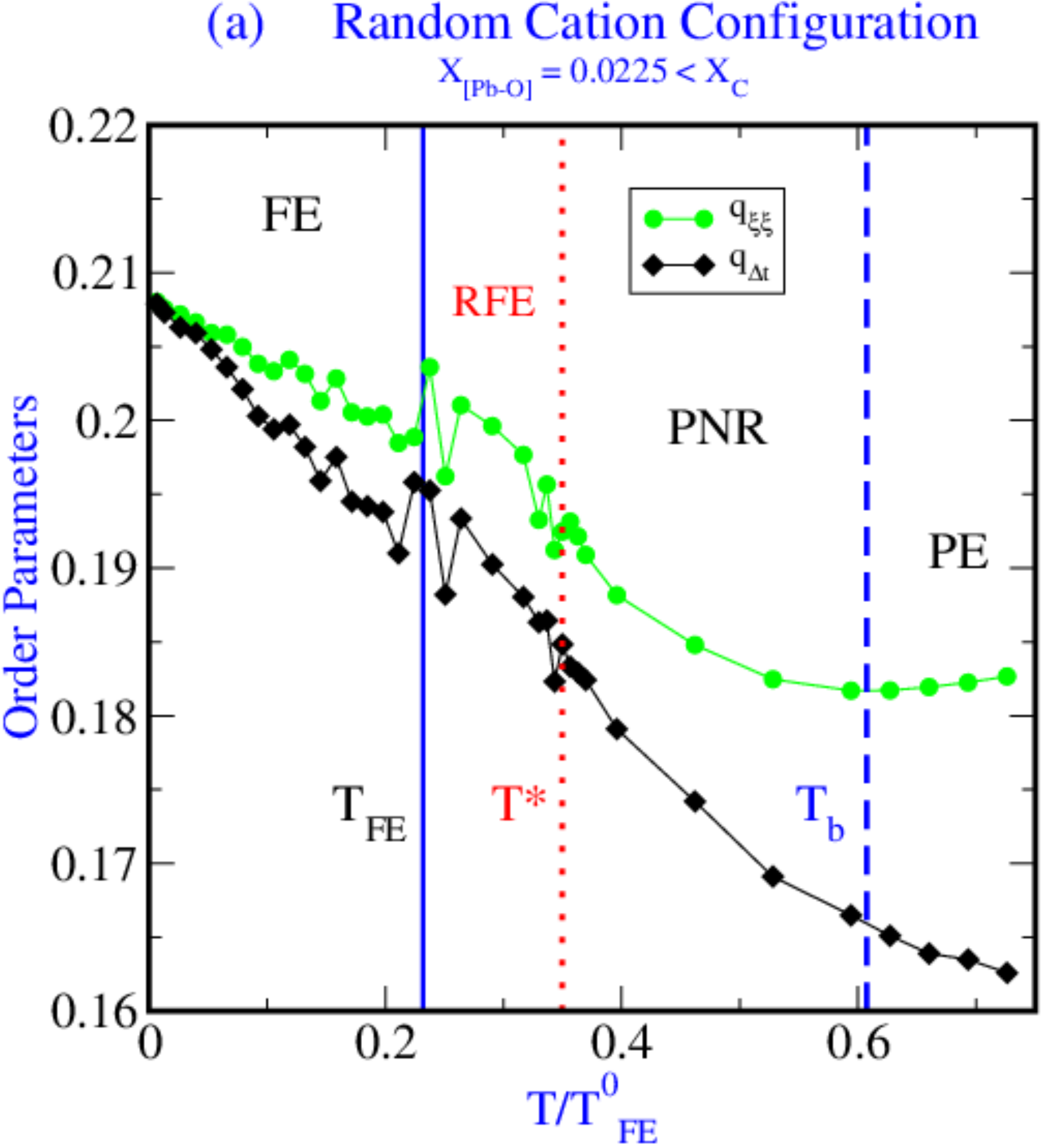}
\hspace{0.5cm}
\includegraphics[width=80mm]{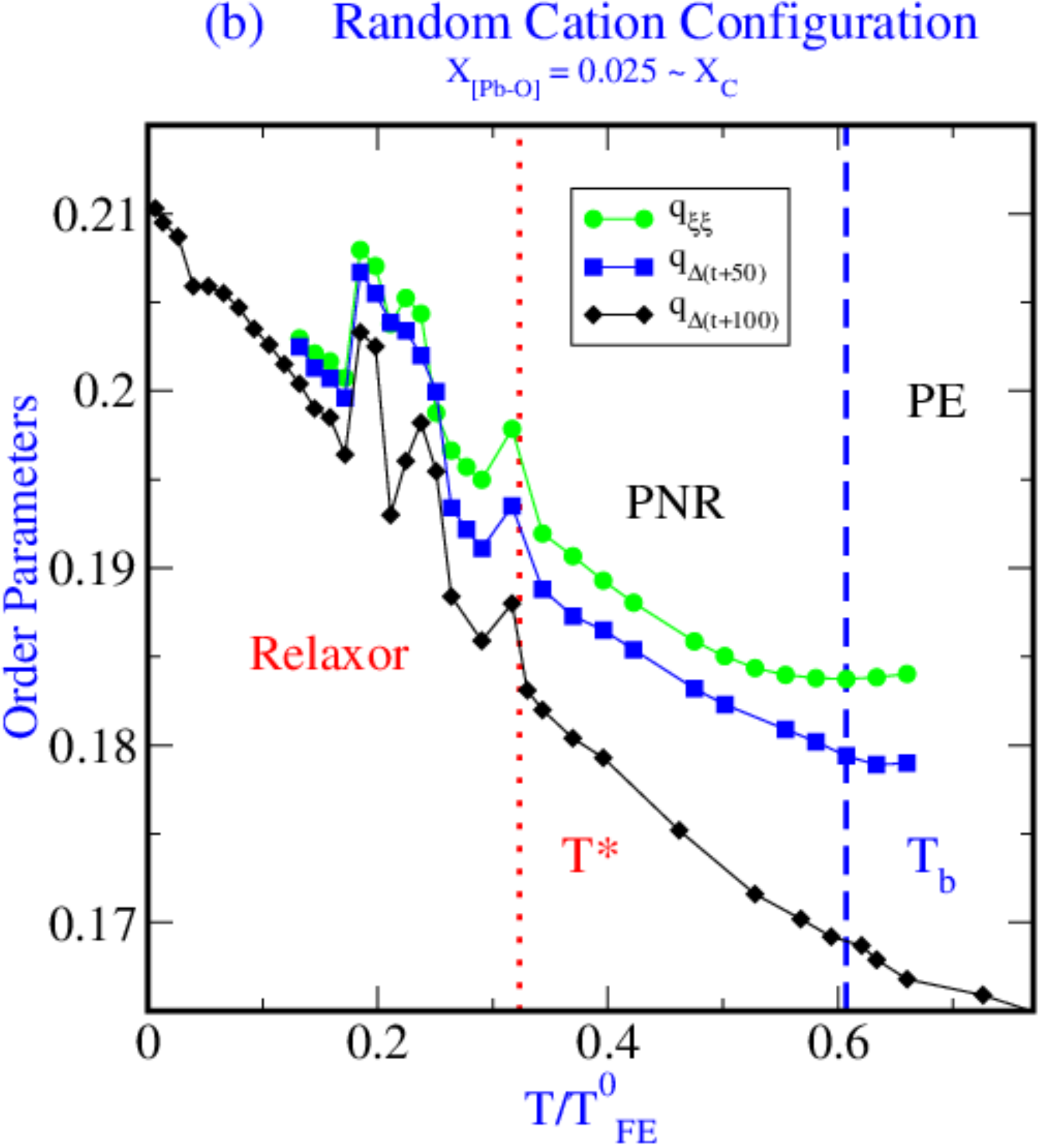}
\vspace{+1.0cm}
\caption{Order parameters as functions of temperature for
$Pb_{1-X}(Sc_{1/2}Nb_{1/2})O_{3-X}$, with a random 
Sc:Nb-cation configuration.
$q_{\xi\xi}(T)$~ and $q_{\Delta t}(T)$~
(defined in Eqns. \ref{xixi:eq} and \ref{QDt:eq}): 
(a) $X_{\rm [Pb-O]}=0.0225 < X_{C}$~ where there is a relaxor ferroelectric
(RFE) with an FE-ground-state; (b) $X_{\rm [Pb-O]}=0.025 \approx X_{C}$~
has no FE-ground-state.  In both (a) and (b), $T^{\bigstar}$~ appears
to mark a weakly first-order transition 
(an $\approx 3\%$~ discontinuity). In (b) $q_{\xi\xi}(T)$, $q_{t+50}(T)$, 
and $q_{t+100}(T)$~ exhibit only small quantitative differences.
}
\label{SSRAND:fig}
\end{figure}

\begin{figure} [!htbp]
\includegraphics[width=75mm]{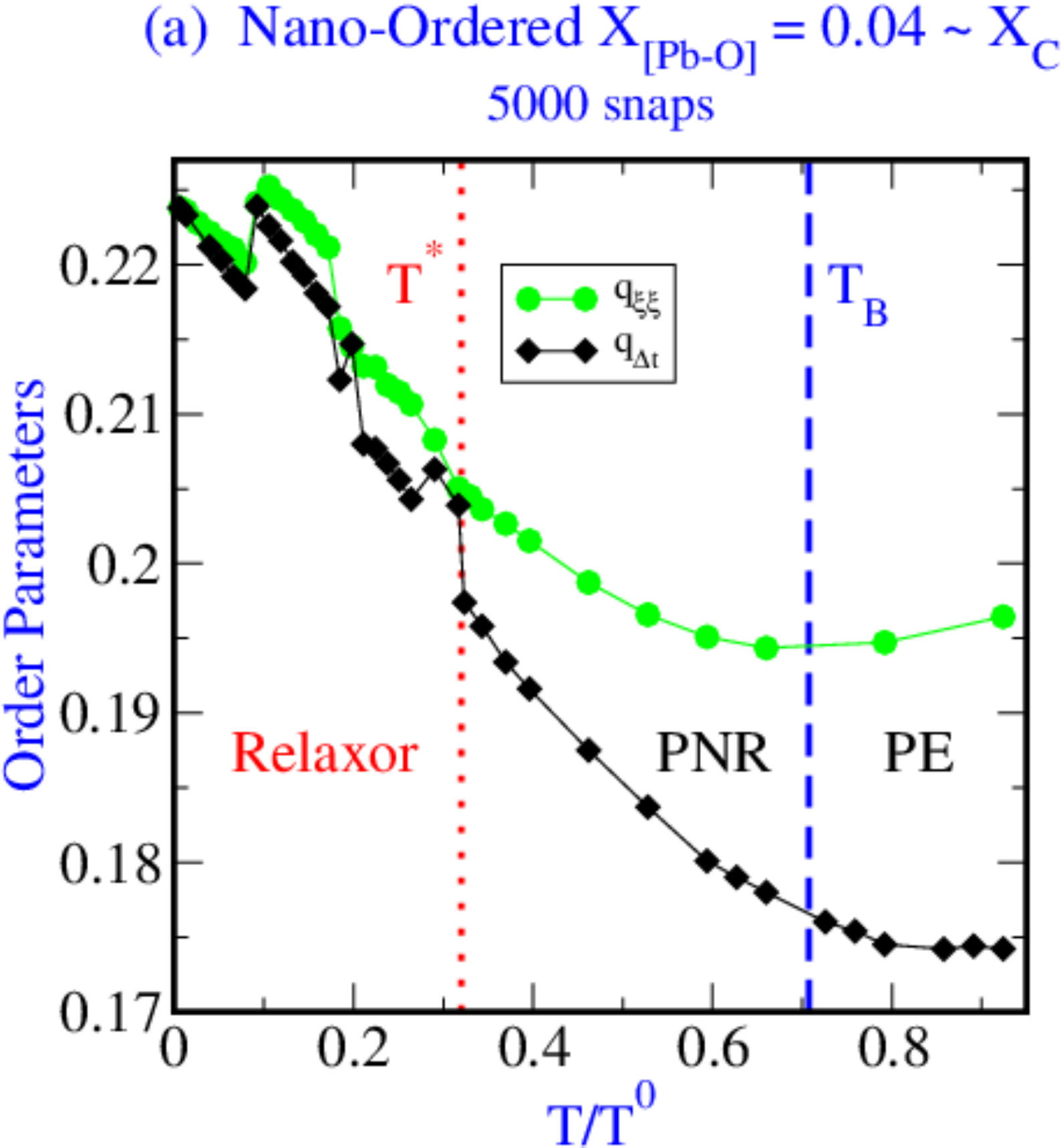}
\hspace{0.5cm}
\includegraphics[width=91mm]{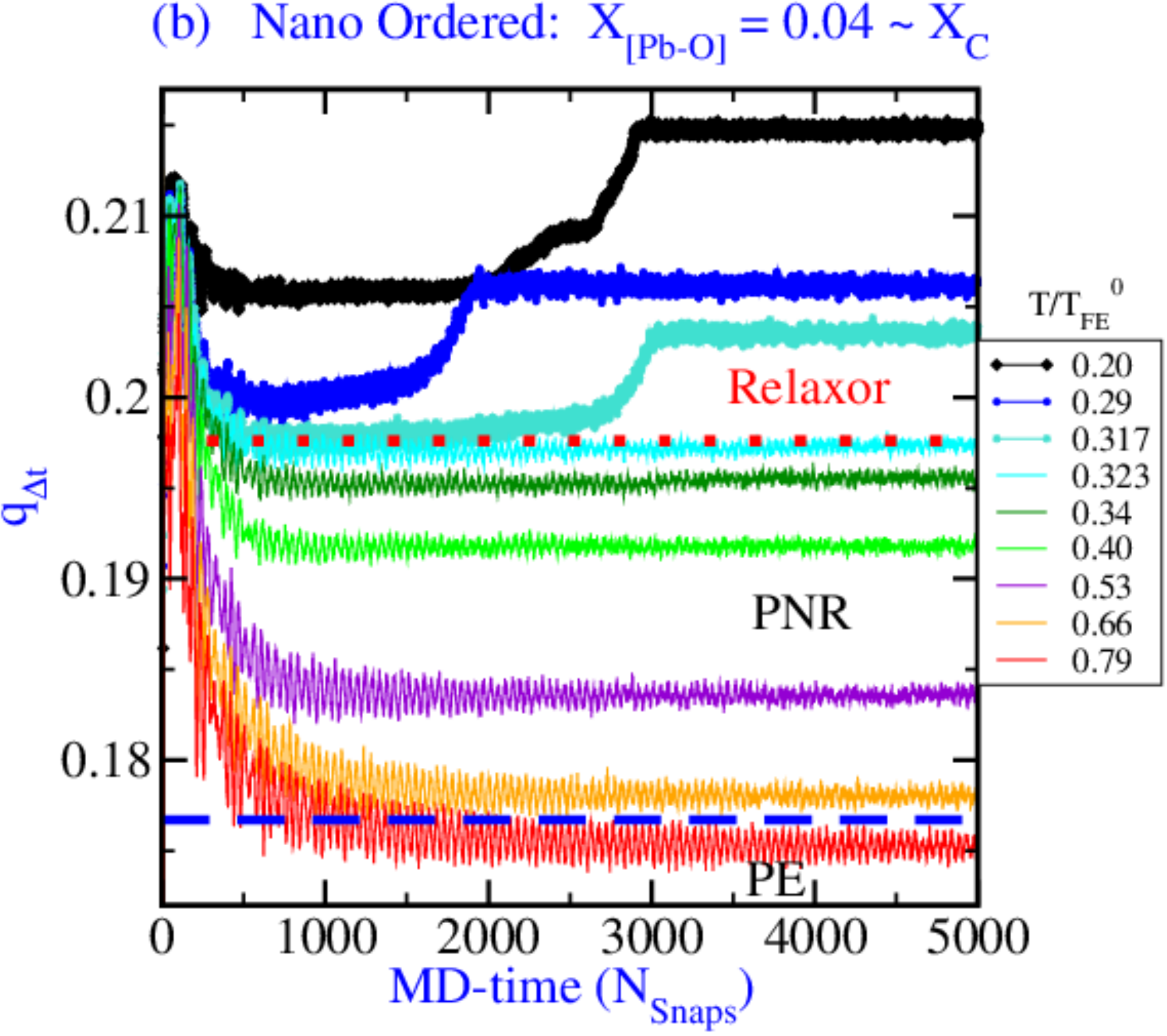}
\caption{Order parameters as functions of temperature for 
a nano-ordered Sc:Nb-cation configuration of 
$Pb_{1-X}(Sc_{1/2}Nb_{1/2})O_{3-X}$, with 25\% ordered
regions in a random matrix: 
(a) $T^{\bigstar}$~ appears to mark a weakly first-order phase transition
(an $\approx 1\%$~discontinuity in $q_{\xi\xi}$~
or $\approx 2\%$~ in $q_{\Delta t}$);
(b) is a plot of $q_{\Delta t}$~
as a function of time, where N$_{snaps}$~ is the number of snapshots in
a 5000 snapshot series. At $T < T^{\bigstar}$, 
above the horizontal dotted line (red online), the 
system traverses local minima before converging.
}
\label{SSNO:fig}
\end{figure}

\begin{figure} [!t]
\includegraphics[width=80mm]{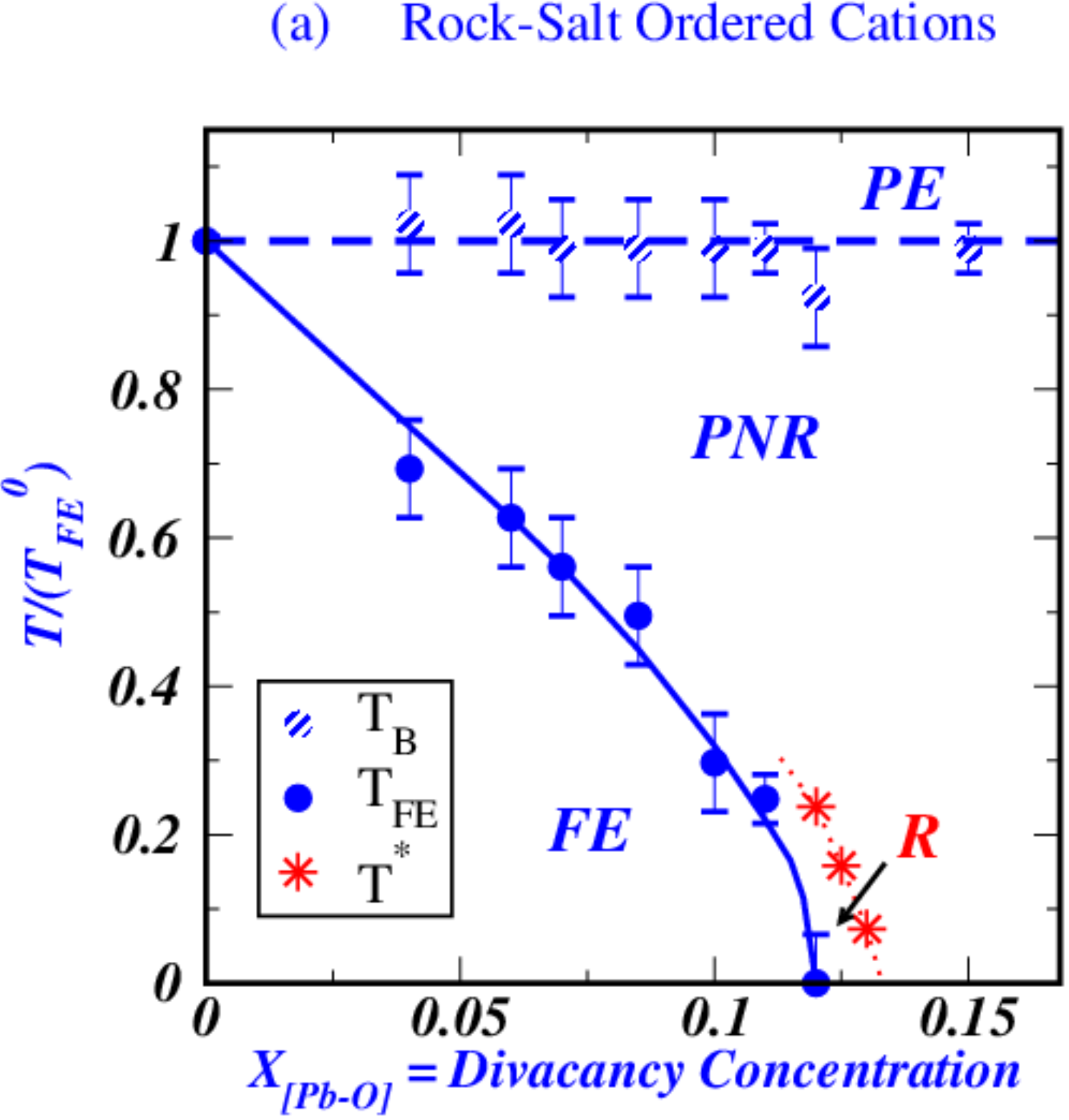}
\hspace{0.5cm}
\includegraphics[width=80mm]{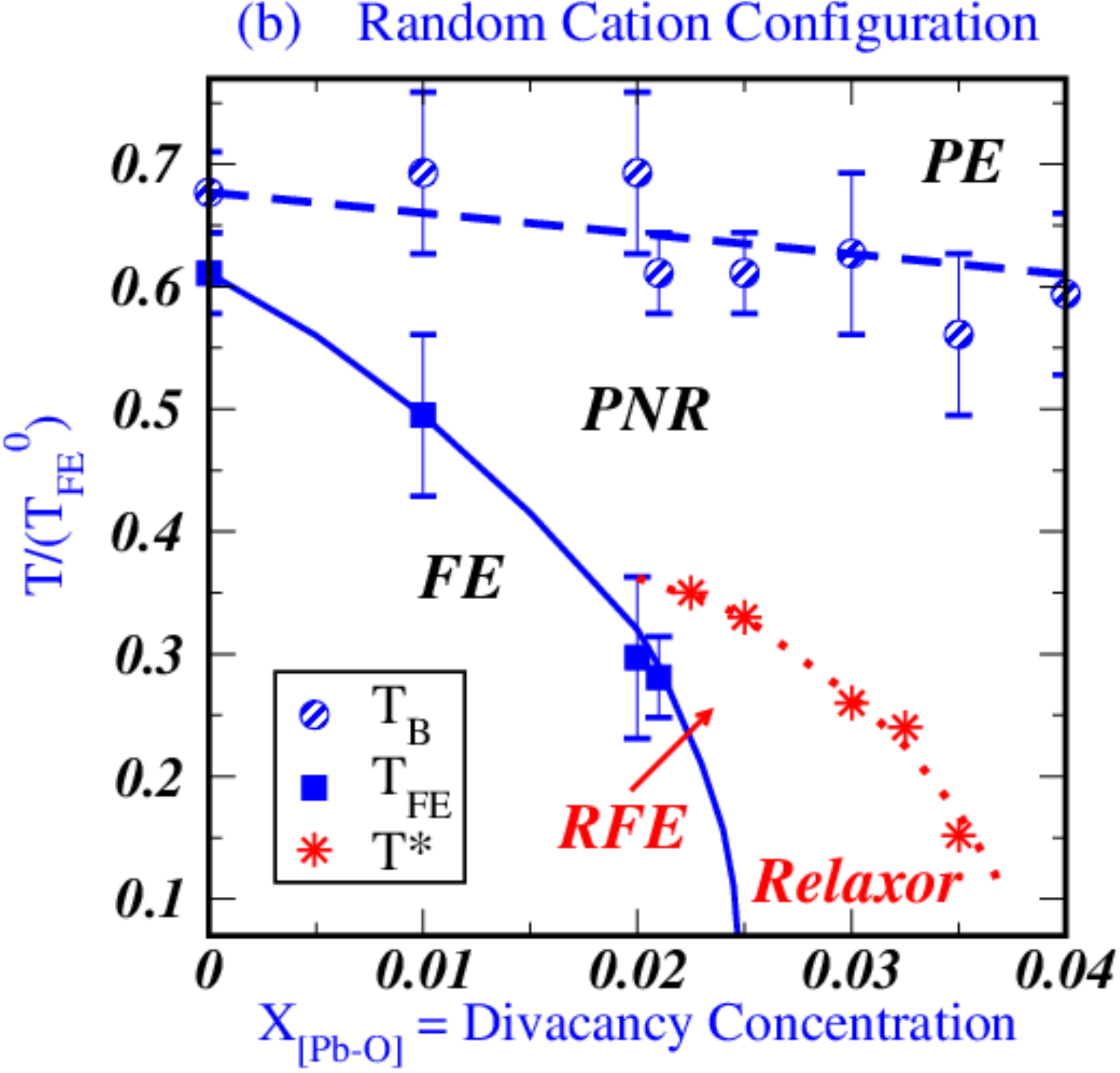}\
\vspace{2.0cm}
\includegraphics[width=80mm]{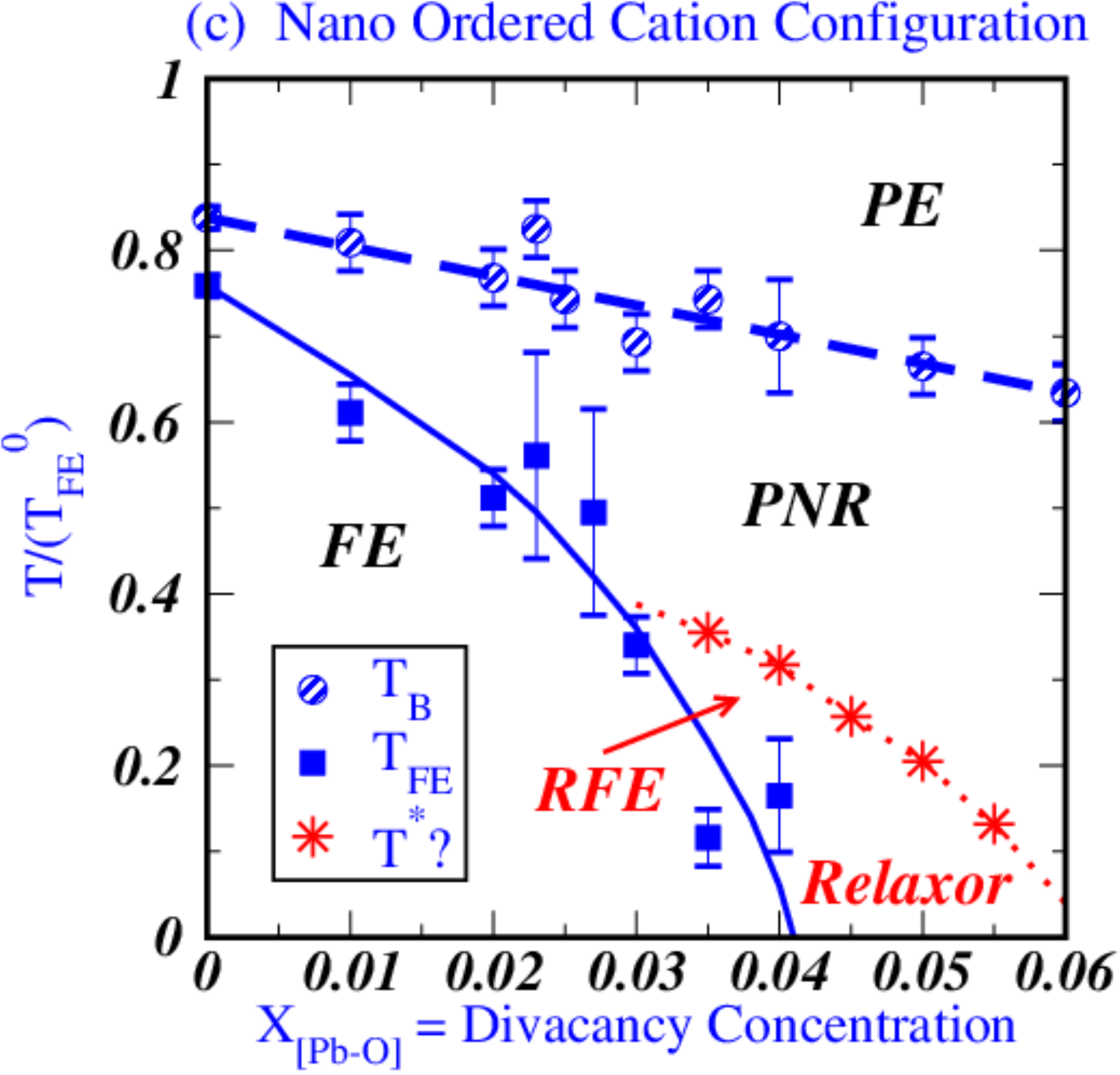}
\vspace{-1.5cm}
\caption{Calculated $X_{\rm [Pb-O]}~vs.~T$~ phase diagrams for the 
system $Pb_{1-X}(Sc_{1/2}Nb_{1/2})O_{3-X}$, with:
a) ideal rock-salt type Sc:Nb-chemical order; 
b) a random Sc:Nb-cation configuration;
c) a nano-ordered Sc:Nb-cation configuration, with ordered
regions in a random matrix (25\% ordered regions that
are $\approx$2~nm in diameter). 
Labels: PE indicates a normal paraelectric; 
PNR indicates a system in which chemically ordered regions, 
with few $\vec{h_{i}}$, have higher polarization than 
the random matrix; FE indicates a ferroelectric ground-state; 
RFE indicates a relaxor-region above the FE-ground-state. 
Dashed lines (blue online) indicate Burns temperatures (T$_B$).
Solid lines (blue online) indicate FE$\rightleftharpoons$PNR,
or FE$\rightleftharpoons$RFE transitions. Dotted lines with
large asterisk-symbols (red online) 
indicate RFE$\rightleftharpoons$PNR or
relaxor$\rightleftharpoons$PNR transitions (crossovers).
}
\label{PD:fig}
\end{figure}

Representative results for order-parameter vs. T curves are plotted
in Figs. \ref{SSORD:fig}, \ref{SSRAND:fig}, and \ref{SSNO:fig}. 
Corresponding phase diagrams are plotted in Figs. \ref{PD:fig}.
In all these plots, T is normalized by T$^0_{FE}$,
the ferroelectric transition temperature ($T_{FE}$) 
of pure ideally rock-salt-ordered
$Pb(Sc_{1/2}Nb_{1/2})O_{3}$. Vertical lines in Figs. \ref{SSORD:fig}, 
\ref{SSRAND:fig}, and \ref{SSNO:fig} indicate 
previously determined \cite{Burton2008} values for 
$T_{FE}$~ and $T_{B}$.
In all these Figures: T$_{FE}$~ is plotted as a solid line (blue 
online); T$_{B}$~ is plotted as a dashed line (blue online); and 
$T^{\bigstar}$~ is plotted as a dotted lines (red online). 
In Figs. \ref{PD:fig}, large asterisk-symbols indicate points 
at which $T^{\bigstar}$~ was located in $q_{\xi\xi}(T)$- 
and $q_{\Delta t}(T)$-curves.

With decreasing T, $q_{\xi\xi}(T)$~ and $q_{\Delta t}(T)$~ typically exhibit: 
broad minima at or near T$_B$; smooth monotonic 
increase in the PNR-region between $T^{\bigstar}$~ and T$_{B}$; 
and erratic increase in the relaxor-region below $T^{\bigstar}$.
The erratic characters of $q_{\xi\xi}(T)$- and $q_{\Delta t}(T)$-curves
in the relaxor-regions of random- and NO-cation-configurations 
are interpreted as indicating glassy behavior. In particular, 
Figs \ref{SSNO:fig}b, which shows the MD time-dependence of 
$q_{\Delta t}(T)$, indicates that in the PNR-region above $T^{\bigstar}$~
$q_{\Delta t}(T)$~ evolves monotonically, however,
in the relaxor-region below $T^{\bigstar}$, $q_{\Delta t}(T)$~
passes through local minima before finding what we take to be 
its final value; as one expects for a glassy material.

\pagebreak

\subsection{Ideal Rock-Salt Chemical Order}

Unlike the random- and nano-ordered cation configurations, the
PNR$\rightleftharpoons$relaxor $transition$~ is subtle
in the ideally NaCl-ordered system; in which [Pb-O]-divacancies are
the only source of $random~ fields$, Figs. \ref{SSORD:fig}.
All three curves in Figs. \ref{SSORD:fig} exhibit changes 
in slope at about $T^{\bigstar}=T/T^0_{FE}\approx0.22$,
but these changes are smaller and less well defined than 
those in Figs. \ref{SSRAND:fig} and \ref{SSNO:fig}; suggesting
that $T^{\bigstar}$~ may mark either a continuous 
PNR$\rightleftharpoons$relaxor transition, or a crossover. 
Also, the erratic variations of order parameters, below $T^{\bigstar}$~ 
that are evident in Figs. \ref{SSRAND:fig} and \ref{SSNO:fig}, 
are either undetectable within MD-precision, or absent in 
the NaCl-ordered system. 

\begin{figure} [!htbp]
\includegraphics[width=160mm]{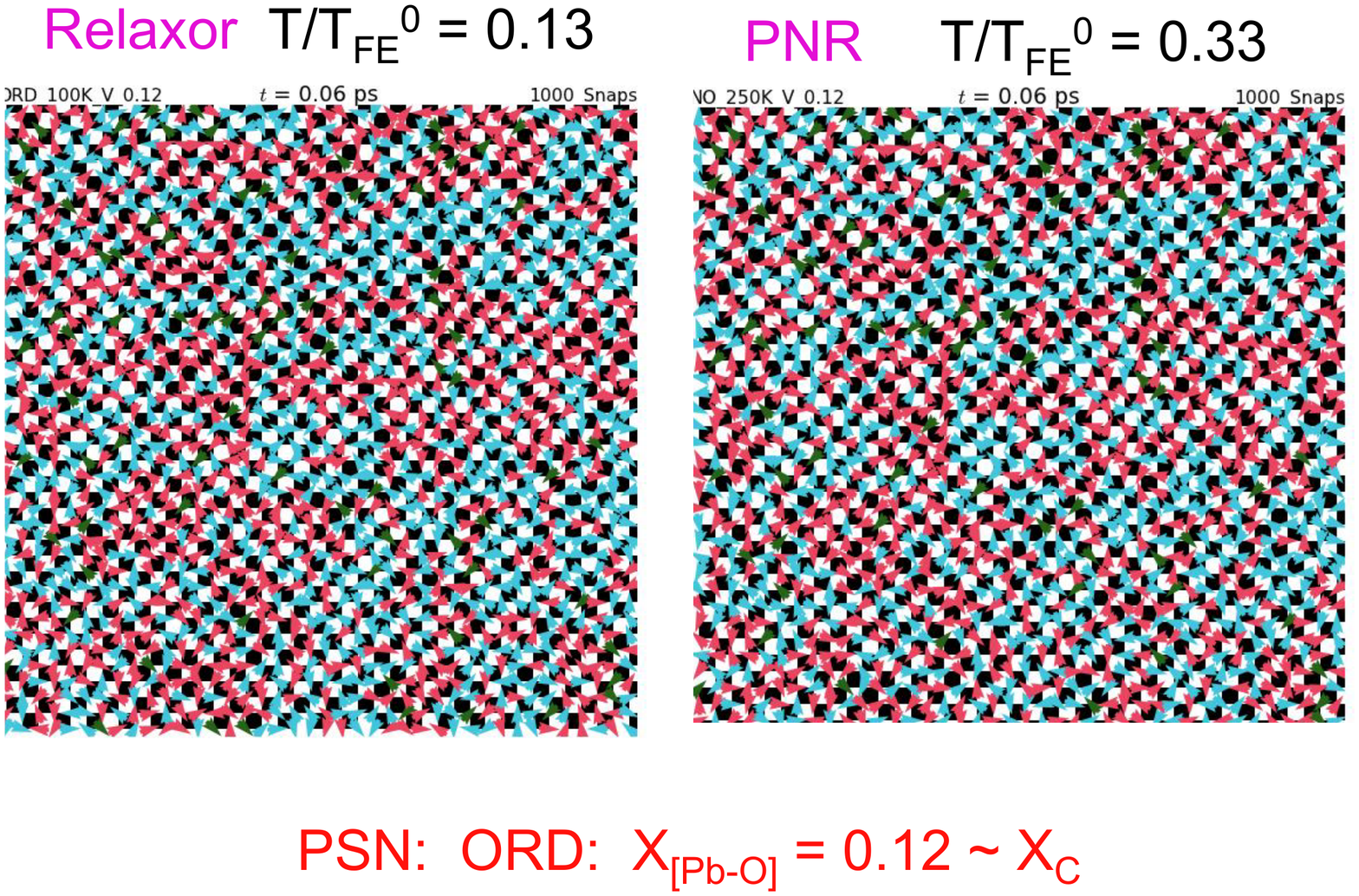}
\vspace{-1.0cm}
\caption{MD-snapshots of the rock-salt ordered cation configuration
black and white squares in the background indicate Sc- and Nb-ions,
respectively. Pb-displacements in the relaxor- (left-panel) and 
PNR-regions (right panel) are plotted as fixed-length arrows:
darker arrows (red online) indicate +z Pb-displacements (out of figure-plane); 
lighter arrows (blue online) indicate -z Pb-displacements (into figure-plane); 
large $<110>$-arrows (dark green online) indicate nn-$[Pb-O]$-divacancies. 
Both panels represent
Pb-displacements after 1000 MD snapshots. Note how similar the 
+z/-z-configurations are in the two panels. This reflects the strong
correlation between the [Pb-O]-divacancy configuration and the 
polar-microstructure over a wide T-range.
}

\label{ORD_micro:fig}
\end{figure}

The rock-salt ordered relaxor has a very different microstructure 
Fig. \ref{ORD_micro:fig} 
than the random- or nano-ordered cation configurations.    
In Fig. \ref{ORD_micro:fig} Pb-displacement patterns and [Pb-O]-divacancy 
configurations are strongly correlated. Hence, even though the
two panels represent a relaxor-, and a PNR-state that is close to T$_B$, 
their Pb-displacement patterns are strikingly similar; reflecting 
the pinning of Pb-displacement patterns to the [Pb-O]-divacancy 
configuration.  Note that the polar microstructure of the rock-salt ordered
system looks more like inter-penetrating, and percolating, +z and -z domains
(out- and into the plane of the figure, respectively), than
like ordered domains in a disordered matrix. 

\begin{figure} [!htbp]
\includegraphics[width=160mm]{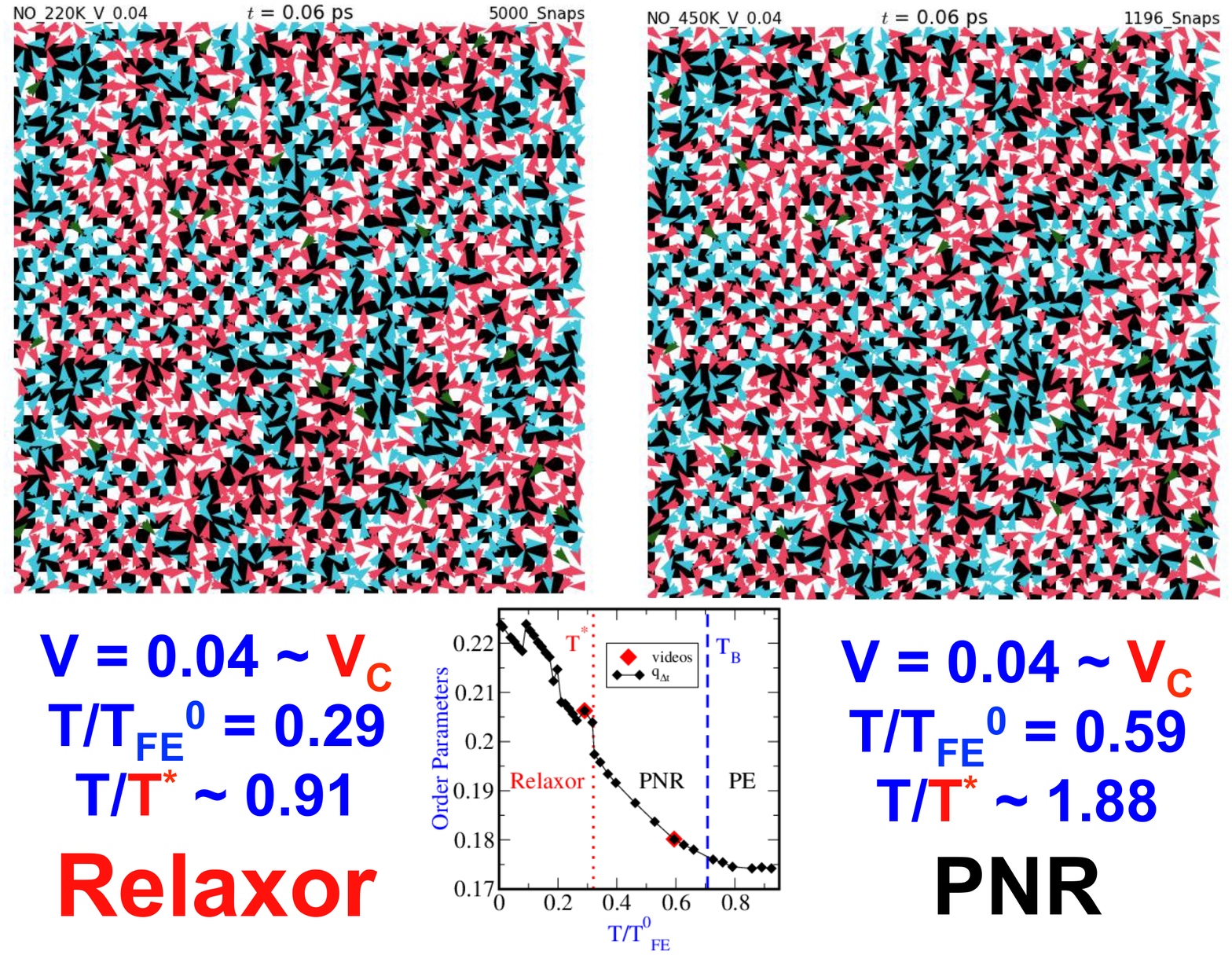}
\vspace{-1.0cm}
\caption{MD-snapshots of the nano-ordered cation configuration after
5000 MD-snapshots, 50,000 MD time-steps: relaxor state is the 
left-panel, and PNR-state near-$T_B$~ is the right panel.  
Squares and arrows that describe 
the Sc:Nb-cation configuration and the Pb-displacements, respectively, 
are as in Fig. \ref{ORD_micro:fig}. Here chemically ordered- and 
disordered regions are evident, and there is a clear positive 
correlation between chemical- and polar-order. Also, as in the 
rock-salt ordered configuration, the +z/-z-configurations are 
strikingly similar, even though the relaxor-
and PNR- snapshots are from very different temperatures 
(large diamonds in central panel, red online).
}

\label{NO_micro:fig}
\end{figure}

\subsection{Random Chemical Disorder and the Nano-Ordered Configuration}

Results for the random- and nano-ordered configurations exhibit very
similar systematics for the $q_{\xi\xi}(T)$- and $q_{\Delta t}(T)$-curves
with decreasing temperature: 
near $T_{B}$, there is a typically a broad minimum; between 
$T_{B}$~ and $T^{\bigstar}$, they increase smoothly and monotonically; 
at $T^{\bigstar}$, there appears to be a (weakly) first-order 
transition, Figs. \ref{SSRAND:fig} and \ref{SSNO:fig}; and below
$T^{\bigstar}$, they vary erratically, and $q_{\Delta t}(T)$~ evolves 
through local minima, Fig. \ref{SSNO:fig}b, before apparently converging. 
Also, there are strong correlations between chemical- and polar-order, 
Fig. \ref{NO_micro:fig} as reported in Burton $et~al.$ \cite{Burton2008}. 

\pagebreak

\section{Discussion}

\subsection{Phase Diagram Topology}

\begin{figure} [!htbp]
\vspace{1.5cm}
\includegraphics[width=80mm]{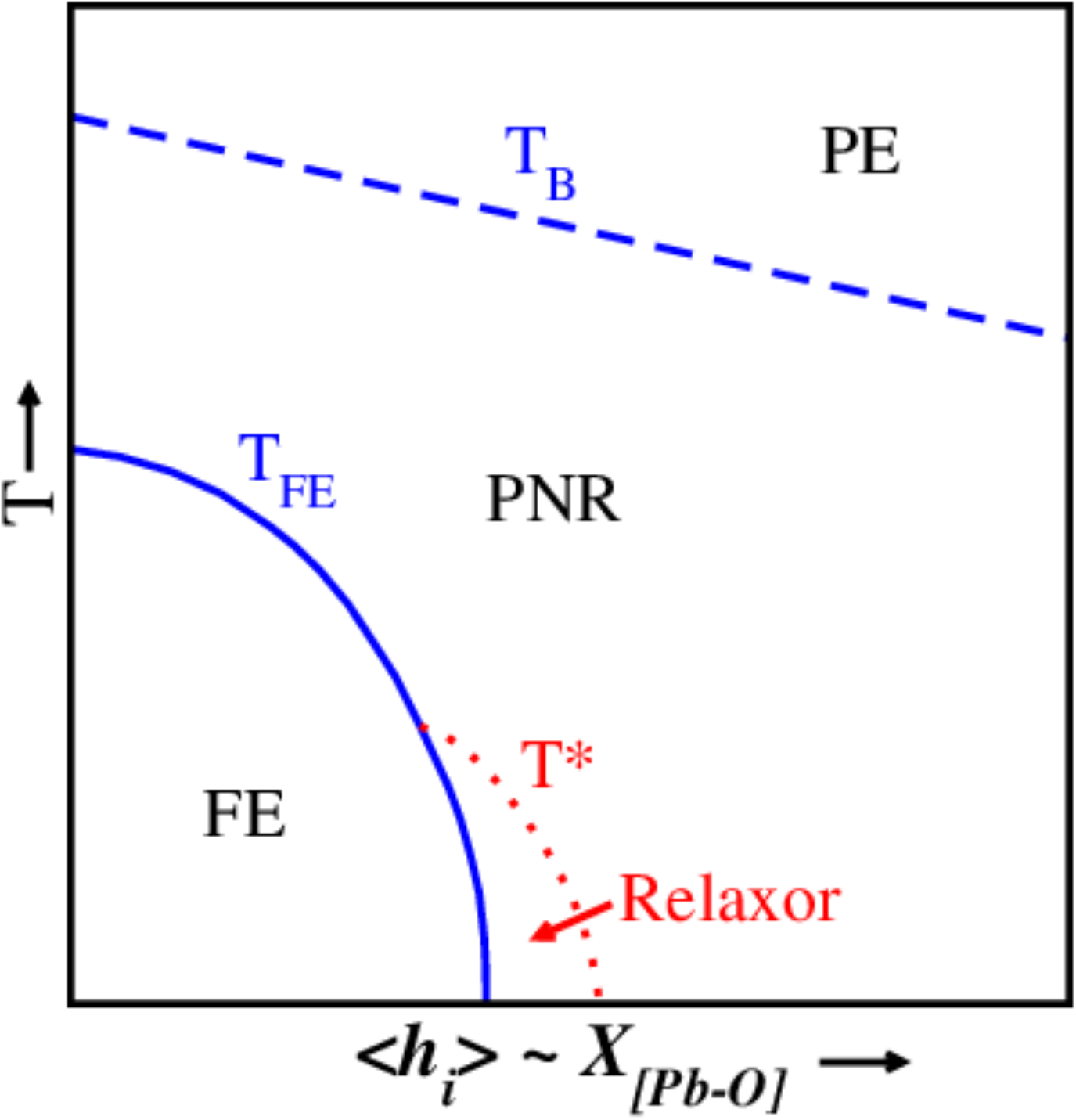}
\vspace{+1.0cm}
\caption{Schematic prototype $\langle \vec{h_i} \rangle $~vs.~T phase diagram for 
$Pb_{1-X}(B,B^{\prime})O_{3-X}$~ relaxor systems. 
}
\label{Proto:fig}
\end{figure}

Notwithstanding the differences between $q_{\xi\xi}(T)$~ and 
$q_{\Delta t}(T)$-curves for the NaCl-ordered configuration vs.
those for the random- and NO-ordered configurations, all three phase diagrams
exhibit the same topology, Figs. \ref{PD:fig}. Given that 
$X_{\rm [Pb-O]}$~ and $\langle \vec{h_i} \rangle $~ are 
interchangeable variables, the phase diagram topology exhibited
in Figs. \ref{PD:fig} can be taken as a prototype for 
$Pb(B,B^{\prime})O_{3}$~ relaxor systems; as depicted in Fig.\ref{Proto:fig}.
In Figs. \ref{PD:fig}, the RFE/relaxor field only occupies a narrow  
$X_{\rm [Pb-O]}$-range from about $X_{C} - 0.015$~ to about $X_{C} + 0.025$; 
$i.e.$~ a limited range of average $\langle \vec{h_i} \rangle $-strength.

\subsection{Comparison With Experiment}

Given the approximations in this model, we do not expect 
$quantitative$~ accuracy in the calculated phase diagrams,
but our results for a random cation configuration (Fig. \ref{PD:fig}b) 
agree reasonably well with experimental data of Chu et al. \cite{Chu0}.
Their dielectric constant measurements of $\epsilon ^{\prime}(T)$~ and
$\epsilon ^{\prime \prime}(T)$~ for almost stoichiometric
PSN [$Pb_{0.998}(Sc_{1/2}Nb_{1/2})O_{2.998}$], and for PSN with
$X_{[Pb-O]} = 0.017 \pm 0.003$~ [$Pb_{0.983}(Sc_{1/2}Nb_{1/2})O_{2.983}$], 
respectively, indicate that the former exhibits a first order 
$PNR\rightleftharpoons FE$~ phase transition, while the latter, 
$Pb_{0.983}(Sc_{1/2}Nb_{1/2})O_{2.983}$, appears to exhibit fully 
relaxor behavior without an FE ground state.  
From Fig. \ref{PD:fig}b one correctly predicts the $PNR\rightleftharpoons FE$~ 
phase transition in the $Pb_{0.998}(Sc_{1/2}Nb_{1/2})O_{2.998}$-sample, but 
one would expect the $Pb_{0.983}(Sc_{1/2}Nb_{1/2})O_{2.983}$-sample to
also have a FE-ground-state, with an intermediate RFE-phase.
In Fig. \ref{PD:fig}b, the calculated critical composition, beyond which there 
is no FE-ground state, is $X_{C} \approx 0.024$. This is at least 
half a percent larger than $Pb_{0.983}(Sc_{1/2}Nb_{1/2})O_{2.983}$~
(the apparent $maximum$~ experimental value), 
which suggests that our model systematically underestimates the strength 
of random fields from charge disorder, vacancies, or both.

\subsection{The PNR$\rightleftharpoons$relaxor transition and criticality}

The apparent predictions of weakly
first-order PNR$\rightleftharpoons$relaxor transitions in the random- 
and nano-ordered cation configurations has an important implication for 
relaxors. Specifically, a weakly first-order transition implies proximity 
to a critical point, and this suggests a simple explanation for the 
extraordinary electro-mechanical properties that are observed in relaxors; 
$i.e.$~ these properties diverge at a critical point,
and are significantly enhanced close to a critical point. 
Indeed, Kutnjak $et~al.$, attributed the giant electromechanical response in 
PMN-PT to a liquid-vapor like critical point. \cite{Kutnjak2006} The
results reported here suggest that the PNR$\rightleftharpoons$relaxor 
transition is $typically$~ close to a critical point; e.g. $close$, 
in the sense that the application of a modest electrical field can drive
the system from weakly first-order to critical.

\subsection{Additional Phase Transitions?}

The experimental phase diagram for the $Eu_XSr_{1-X}S$~
exhibits a ferromagnetic$\rightleftharpoons$SG transition,\cite{Maletta1979} and in
$Fe_{1-X}Au_{X}$~ there are ferromagnetic$\rightleftharpoons$Mixed-phase-
and SG$\rightleftharpoons$Mixed-phase-transitions \cite{Coles1978};
in which, the Mixed-phase is ferromagnetic but replica-symmetry breaking (RSB)
\cite{Castellani2005}.
Compelling evidence of analogous transitions was not detected in this work 
\cite{LowT}, but such transitions are not ruled out, and there is clear similarity between 
relaxor- and magnetic spin-glass phase diagrams: Fig. \ref{Proto:fig} and Table I.  
  
\begin{table}
\begin{center}
\caption{Relaxor vs. Magnetic Spin-Glass Analogy.}
\end{center} 
\begin{tabular}{|c|c|} \hline
{$Relaxor$}                         &  {$Magnetic~Spin-Glass$}   \\ \hline \hline 
{$PE=paraelectric$ }                &  {$PM=paramagnetic$}       \\ \hline
{$PNR=Polar~Nano~Regions$}          &  {$SPM=superparamagnetic$} \\ \hline 
{$FE=Ferroelectric$ }               &  {$FM=Ferromagnetic$}      \\ \hline
{$RSB=Replica-Symmetry-Breaking$}   &  {$RSB$}                   \\ \hline
{$RFE/relaxor$}                     &  {$SG=Spin Glass$}         \\ \hline \hline
\end{tabular}  
\label{RvSG} 
~~~~\\
\vspace{2.0mm}
\end{table}

\section{Conclusions}

The phase diagrams presented in Burton $et~al$. \cite{Burton2008} were
incomplete because they omitted $T^{\bigstar}(X_{\rm [Pb-O]})$-curves;
$i.e.$~ delineation of the RFE/relaxor-phase field.
Results presented here include: calculations of 
$T^{\bigstar}(X_{\rm [Pb-O]})$-curves;  suggest a prototype
relaxor phase diagram topology; and strongly support the analogy
between relaxors and magnetic spin-glasses, with respect to phase diagram
topology, Table I.

The combination of soft-spins with explicit 1'st-3'rd nn-pairwise 
pseudospin-pseudospin interactions, 
4'th-39'th nn dipole-dipole interactions, and random fields, is evidently 
sufficient to model heterovalent $Pb(B,B^{\prime})O_{3}$~ relaxor systems. 
Both the self-orvelap order parameter and the autocorrelation function
appear to be good order parameters for locating 
$T^{\bigstar}(X_{\rm [Pb-O]})$-curves, and for demonstrating the glassy 
character of the relaxor-phase, which only occupies a narrow range in
$X_{\rm [Pb-O]}$, or equivalently in $\langle \vec{h_i} \rangle $.

Previous conclusions \cite{Burton2005,Burton2008} about the strong 
correlation between chemically ordered regions and PNR are reinforced, 
with the addition that the orientations of PNRs become more static in 
the relaxor region, below the PNR$\rightleftharpoons$relaxor transition.  
In the random- and nano-ordered cation configurations $T^{\bigstar}$~ appears
to be a weakly first-order transition, but results for the rock-salt 
ordered configuration are suggestive of a continuous transition
or a crossover. Hence, chemical inhomogeneities such as chemical 
short-range order apparently amplify relaxor character.

\section{Acknowledgements}

B.P. Burton thanks D. Sherrington for inspiring this work and for
helpful discussions.


\begin{thebibliography}{999}
\bibitem{QUENCH}
The phrase "quenched chemical disorder" implies that the chemical 
configuration is fixed within the temperature range that polar 
order-disorder is analyzed.
\bibitem{LowT}
The strongest suggestion of such a transition is in the NO configuration
where $q_{\Delta t}(t)$~ exhibits a low-T reduction
to a plateau, which does not seem to define a consistent phase-boundary
or crossover (e.g. Fig. supplimentary material).
\end{thebibliography}
\end{document}



\section{SUPPLEMENTARY MATERIAL}

\begin{figure}[!htbp]
\begin{center}
\vspace{0.10in}
\includegraphics[width=10.0cm,angle=0]{NOPD_LT.pdf}
\end{center}
\caption{An enlargement of the relaxor region in Fig. 2b in which
large black asterix-symbols indicate points of $possible$~ low-T
transition. Because similar analyses of the NaCl-ordered and random cation
configurations gave no indication of plausible phase (crossover)
boundaries, we do not consider these results convincing.
}
\label{fg:NOPD_LT}
\end{figure}